\documentclass[twocolumn,prb,showkeys,showpacs,superscriptaddress]{revtex4-1}
\usepackage[T1]{fontenc}
\usepackage[latin9]{inputenc}
\usepackage{units}
\usepackage{graphicx}

\makeatletter

\providecommand{\tabularnewline}{\\}

\usepackage{dcolumn}
\usepackage{bm}
\usepackage{times}
\usepackage{graphics}
\usepackage{color}

\makeatother

\begin{document}
\title{Emergence of a metallic meta-stable phase induced by electrical current
in Ca$_{2}$RuO$_{4}$}
\author{C.~Cirillo}
\affiliation{CNR-SPIN, c/o Università degli Studi di Salerno, I-84084 Fisciano
(Sa), Italy}
\affiliation{Dipartimento di Fisica ``E.R. Caianiello'', Università degli Studi
di Salerno, I-84084 Fisciano (Sa), Italy}
\author{V.~Granata}
\affiliation{Dipartimento di Fisica ``E.R. Caianiello'', Università degli Studi
di Salerno, I-84084 Fisciano (Sa), Italy}
\affiliation{CNR-SPIN, c/o Università degli Studi di Salerno, I-84084 Fisciano
(Sa), Italy}
\author{G.~Avallone}
\affiliation{Dipartimento di Fisica ``E.R. Caianiello'', Università degli Studi
di Salerno, I-84084 Fisciano (Sa), Italy}
\affiliation{CNR-SPIN, c/o Università degli Studi di Salerno, I-84084 Fisciano
(Sa), Italy}
\author{R.~Fittipaldi}
\affiliation{CNR-SPIN, c/o Università degli Studi di Salerno, I-84084 Fisciano
(Sa), Italy}
\affiliation{Dipartimento di Fisica ``E.R. Caianiello'', Università degli Studi
di Salerno, I-84084 Fisciano (Sa), Italy}
\author{C.~Attanasio}
\affiliation{Dipartimento di Fisica ``E.R. Caianiello'', Università degli Studi
di Salerno, I-84084 Fisciano (Sa), Italy}
\affiliation{CNR-SPIN, c/o Università degli Studi di Salerno, I-84084 Fisciano
(Sa), Italy}
\author{A.~Avella}
\affiliation{Dipartimento di Fisica ``E.R. Caianiello'', Università degli Studi
di Salerno, I-84084 Fisciano (Sa), Italy}
\affiliation{CNR-SPIN, c/o Università degli Studi di Salerno, I-84084 Fisciano
(Sa), Italy}
\author{A.~Vecchione}
\affiliation{CNR-SPIN, c/o Università degli Studi di Salerno, I-84084 Fisciano
(Sa), Italy}
\affiliation{Dipartimento di Fisica ``E.R. Caianiello'', Università degli Studi
di Salerno, I-84084 Fisciano (Sa), Italy}
\begin{abstract}
A comprehensive study of the behavior of the Mott insulator Ca$_{2}$RuO$_{4}$
under electrical current drive is performed by combining two experimental
probes: the macroscopic electrical transport and the microscopic X-Ray
diffraction. The resistivity, $\rho$, vs electric current density,
$J$, and temperature, $T$, $\rho$(J,T), resistivity map is drawn.
In particular, the meta-stable state, induced between the insulating
and the metallic thermodynamic states by current biasing Ca$_{2}$RuO$_{4}$
single crystals, is investigated. Such an analysis, combined with
the study of the resulting RuO$_{6}$ octahedra energy levels, reveals
that a metallic crystal phase emerges in the meta-stable regime. The
peculiar properties of such a phase, coexisting with the well-established
orthorhombic insulating and tetragonal metallic phases, allow to explain
some of the unconventional and puzzling behaviors observed in the
experiments, as a negative differential resistivity. 
\end{abstract}
\maketitle

\section{Introduction}

Ca$_{2}$RuO$_{4}$ (hereafter Ca-214) is a paramagnetic Mott insulator
subject of extensive experimental and theoretical studies \citep{Gorelov_10,Sow_17,Ricco_18,Das_18,Porter_18}.
The richness of its phase diagram \citep{Steffens_05,Sow_17} and
the strong interplay between electronic, structural, magnetic and
orbital degrees of freedom make the full comprehension of the physics
of this system challenging \citep{Mizokawa_01,Gorelov_10,Sutter_17,Das_18}.
This material indeed exhibits very different responses, both in the
magnetic and transport properties, to different combination of temperature
\citep{Nakatsuji_01,Cao_97}, pressure \citep{Steffens_05,Nakamura_07,Alireza_10},
doping \citep{Carlo_12,Ricco_18,Sutter_19}, and electrical field
\citep{Nakamura_13}.

\begin{figure}
\noindent \begin{centering}
\includegraphics[width=8cm]{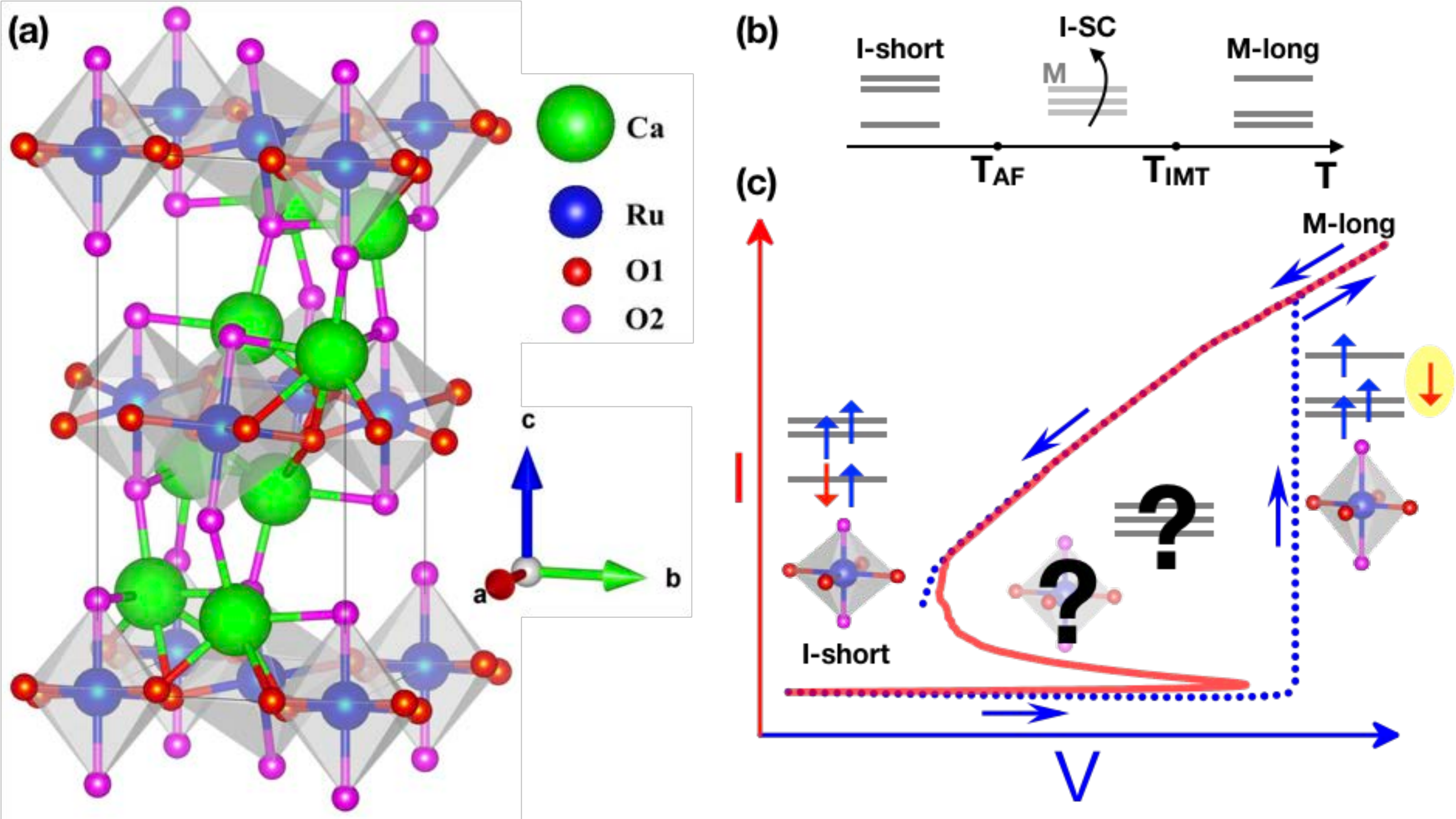}
\par\end{centering}
\noindent \caption{(a) Crystallographic unit cell. (b) Sketch of the temperature evolution
of the $t_{2g}$ orbitals of Ru, see text for details. (c) Cartoon
of the $I-V$ curve: the hysteretic path for an electrical potential
drive (blue) and the behavior for an electrical current drive (red)
are shown. The characteristic octahedron shapes (the axes ratios are
exaggerated for illustration purposes) and levels/electrons characteristic
arrangements are reported for the different regions of the $I-V$
characteristic, corresponding to S-\emph{Pbca} (I-short) and L-\emph{Pbca}
(M-long). The intermediate state with $dI/dV<0$ is the main objective
of the paper.\label{fig1}}
\end{figure}

Ca-214 is a layered perovskite oxide with \emph{Pbca} space-group
symmetry whose crystallographic unit cell contains four formula units
{[}see Fig.~\ref{fig1}(a){]}. The fundamental structural units are
RuO$_{6}$ octahedra arranged in corner-shared planes alternated by
layers containing the Ca atoms. With respect to the ideal tetragonal
structure (with lattice parameters $a=b$, $c$), the octahedra bear
alternating rotations (about the apical Ru-O2 bond; $z$ hereafter),
tilts (of $z$ with respect to the $ab$-plane initially containing
the Ru-O1 bonds; $x$ and $y$ hereafter) and distortions (making
$x$ and $y$ slightly different) \citep{Braden_98} (see Appendix).
The ratios between $x$ and $y$ and, in particular, between their
average, $\bar{x}$, and $z$ determine the relative energies of the
$t_{2g}$ orbitals of Ru ($d_{xy}$, $d_{yz}$, $d_{xz}$), which
are the electrons responsible of transport as well as all other response
properties. $x/y$ rules the relative position of $d_{yz}$ and $d_{xz}$
levels while $z/\bar{x}$ rules the relative position of $d_{xy}$
with respect to the $d_{yz}$-$d_{xz}$ doublet (see Appendix).

As schematically shown in Fig.~\ref{fig1}(b), these ratios change
with the temperature, $T$. In particular, $z/\bar{x}$ increases
with $T$, as $c/\bar{a}$ does ($\bar{a}$ is the average between
$a$ and $b$). For temperatures below $T_{AFM}\approx\unit[110]{K}$
the ratio is lower than $1$. As a consequence, the system is an antiferromagnetic
(AFM) insulator \citep{Nakatsuji_01} with $d_{xy}$ lower than $d_{yz}$-$d_{xz}$
doublet and the four electrons per Ru atom arranged as shown in Fig.~\ref{fig1}(c)
(I-short). At intermediate and ambient $T$, $z/\bar{x}$ goes through
about $1$, which results in a paramagnetic Strongly Correlated Mott
insulator (I-SC) with the three levels almost degenerate (M), before
both $d_{yz}$ and $d_{xz}$ go through a Mott-Hubbard splitting \citep{Gorelov_10,Zhang_17}.
Finally, when $z/\bar{x}$ is sufficiently larger than $1$, above
$T_{IMT}=\unit[357]{K}$, the system undergoes an Insulator-Metal
transition (IMT) \citep{Nakatsuji_01} with the four electrons per
Ru arranged as shown in the M-long configuration reported in Fig.~\ref{fig1}(c).
IMT is accompanied by a crystallographic transition from a tetragonal
(L-\emph{Pbca}, L stands for long c) to an orthorhombic (S-\emph{Pbca},
S $\rightarrow$ short c) phase, so dramatic to break the crystals
into pieces \citep{Nakamura_13}.

The strong link between conduction and structural properties \citep{Gorelov_10,Zhang_17}
paves the way to control the electronic behavior by strain/epitaxial
growth \citep{Kikuzuki_10} or by inducing nonlinear phononic effects,
for instance through intense terahertz radiation \citep{Rini_07,Ehrke_11,Ichikawa_11}.
Another relevant drive to induce the IMT is the electric field, despite
the structural changes indirectly induced in such a case are not yet
clarified. Indeed, the electric field tuning of the conduction regime
is of particular interest, since at room $T$ the metallic state can
be induced by a threshold field of about $E_{th}\approx\unit[50\div100]{V/cm}$
\citep{Nakamura_13,Okazaki_13,Sow_17}, almost three orders of magnitude
lower than in other Mott insulators \citep{Janod_15}. This circumstance
is very promising for possible applications in next-generation oxide
electronics. As in other Mott materials \citep{Imada_98,Janod_15},
the IMT is accompanied by resistivity changes of several orders of
magnitude \citep{Nakamura_13}. As a first order transition, IMT is
generally unveiled by hysteretic electrical transport \citep{Limelette_03,Nakamura_13}
for voltage drive {[}Fig.~\ref{fig1}(b), blue dotted line{]}. Instead,
voltage-current $V-I$ characteristics with negative slope were reported
for dc current drive Fig.~\ref{fig1}(c), red line \citep{Okazaki_13,Zhang_19}.

However, one should be aware that different measurement protocols
exist in the literature under the simple names of voltage or current
drive. The difference in the procedures on one hand gives new perspective
to look at an interesting system such Ca-214, but makes also difficult
to compare the results obtained in different works. Recently, the
investigation of non-equilibrium electronic and crystallographic phases
emerging by current or voltage biasing the crystals gained much attention.
Indeed, a new crystal structure supposing to be the manifestation
of a new semi-metallic state was reported by a measurement protocol
completely different from the one presented in this work \citep{Bertinshaw_18},
while alternating insulating and metallic regions arranged in stripes
patterns at the M-I phase boundary were observed in the regime of
controlled constant current flow \citep{Zhang_19}. Moreover, it is
now accepted that dc current biasing can be used to control the magnetic
properties of the system, since, under current flow, strong diamagnetism
is induced in pure Ca-214 and in Ca$_{3}$Ru$_{1-x}$Ti$_{x}$O$_{7}$
\citep{Sow_17,Sow_19} and AFM order is suppressed in pure Ca-214
\citep{Bertinshaw_18,Sow_19}.

In this work, the electrical response of Ca-214 single crystals is
investigated as a function of both $T$ and the bias-current density,
$J$, in the conduction regimes spanning from the insulating to the
meta-stable (MS) state, precursor of the metallic one, where non-equilibrium
processes possibly take place. In this way, the resistivity map, $\rho(J,T)$,
of the system, where $\rho$ is the electrical resistivity, is built.
This study, systematically performed on a large number of crystals,
is an extremely valuable starting point for further investigations,
since it naturally highlights the different conducting regimes, as
well as the characteristic temperatures and current densities, at
which they set in. In particular, here the attention focuses on the
less explored MS state, since poor information are currently available
concerning both the conduction mechanisms and the corresponding crystallographic
structure. For these reasons, the transport measurements are combined
with X-Ray Diffraction (XRD) spectra acquired as a function of $J$,
at room $T$.

\section{Experimental Methods}

High quality Ca-214 single crystals were grown by floating zone technique
as described in Ref. \citep{Fukazawa_00}. The typical average dimensions
of the analyzed crystals are about $\unit[(2.0\times1.0\times0.150)]{mm^{3}}$.
Great care was paid to the reproducibility of the presented results.
At this purpose a big amount of data was collected on several Ca-214
single crystals, which all behaved consistently. This assures the
reliability of the presented measurements.

The phase diagram of Ca-214 is very rich as well as quite far from
being fully explored and understood. For this reason, an extremely
precise control of the actual state of the sample, as a function of
the external conditions, is required in order efficiently study this
system. Moreover, an absolutely methodical approach is essential to
obtain reproducible and scientifically sound results. In this respect,
it is necessary to clarify that many different measurement protocols
exist in the literature under the simple names of current or voltage
drive. For a system such as Ca-214, with unconventional and very slow
responses to electric drive, this leads to the great opportunity of
having many different and interesting perspectives that all contribute
to the overall understanding of the complex physics of this material.

On the other hand, the comparison of the results obtained by different
experimental procedures may not be easy. Here, a very straightforward
measurement protocol was used, namely the sample was current biased
in a continuous mode, with the use of a steady current source. This
approach can give access to different states of the system compared
with those already reported in the literature. For instance, in the
work of Bertinshaw et al., the authors first use the voltage to bias
the sample, and once the switching to the metallic phase is achieved,
let an electrical current to flow in the system \citep{Bertinshaw_18}.
Instead, in Ref. \citep{Zhang_19}, voltage and current are simultaneously
controlled by the use of two variable resistors.

Here, electrical transport measurements, both resistivity versus temperature,
for different values of the bias current, and $V-I$ characteristics
as a function of $T$, were performed with a two probe method by current
biasing the crystals along the \textit{c}-direction with a Keithley
2635 sourcemeter and reading the voltage drop with a Keithley 2182A
nanovoltmeter. Due to the high resistance values of the crystals compared
to the ones of the wiring and the contacts, this method does not affect
the measurement accuracy \citep{Nakamura_13,Sow_17,Sow_19}. The electrical
current was chosen as the biasing stimulus since it is capable to
drive the system in to an intermediate state which, as demonstrated,
does not have an equilibrium analog and strongly differs from the
insulating or the metallic thermodynamic phases explored by the voltage-driven
measurement. The accessible area of the resistivity map is determined
by the limit of the sourcemeter, which was set at $\unit[200]{V}$.

Extreme attention was paid to adopt all the precaution necessary to
maximize sample cooling as well as to reduce contact resistance at
the sample ends. First, in order to keep contact resistance as low
as possible, silver pads were sputtered on the crystal faces from
which gold wires (diameter 25 $\mu$m) were connected by an epoxy
silver-based glue with the external wiring. Then, in order to achieve
a fair temperature control, the thermal coupling between the sample
and the Cernox thermometer was carefully implemented: the crystals
were thermally anchored with a small amount of cryogenic high vacuum
grease on a custom-built dip probe on the massive high-thermal-conductivity
copper sample holder in which the thermometer was embedded, in close
contact with the crystal. The temperature was changed by lowering
the probe into a cryogenic liquid nitrogen storage dewar by profiting
of the temperature stratification naturally occurring in the vapor
space above the liquid surface. The thermal stability is guaranteed
by the proper design of the copper sample holder and by the extremely
slow temperature sweeps.

\begin{figure*}
\noindent \begin{centering}
\begin{tabular}{ccc}
\includegraphics[width=8cm]{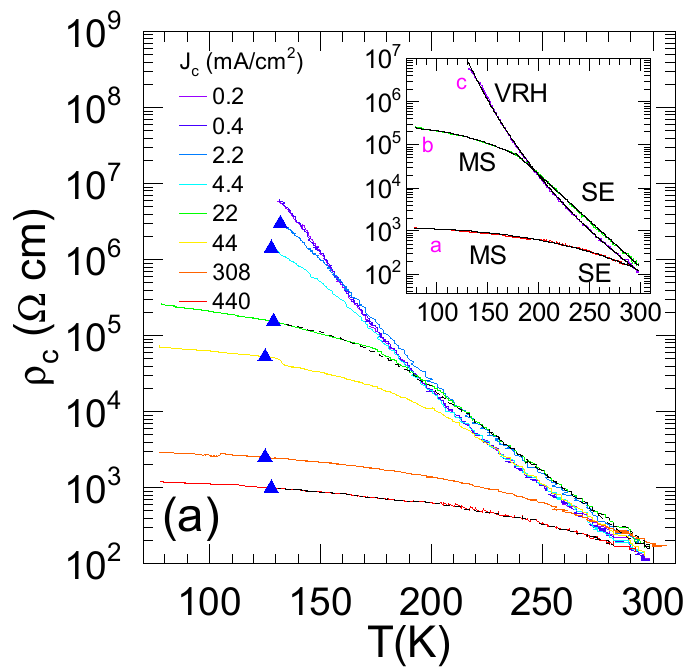} &  & \includegraphics[width=8cm]{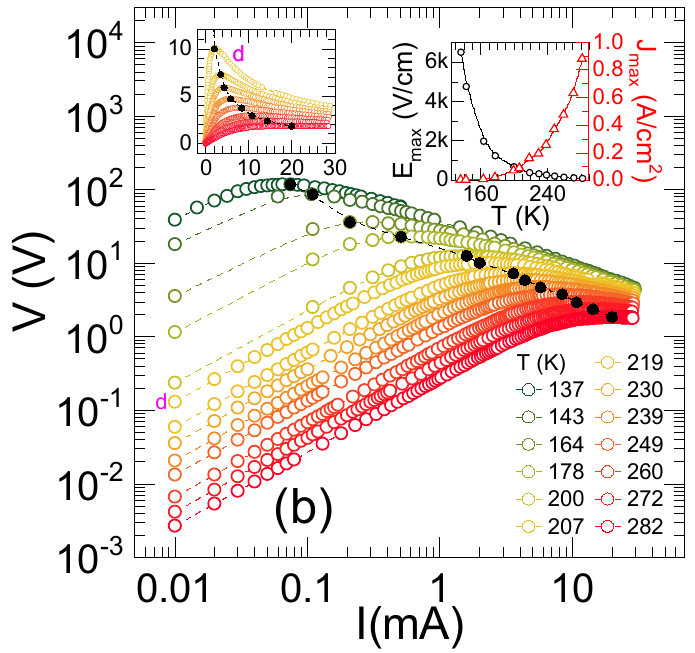}\tabularnewline
\end{tabular}
\par\end{centering}
\caption{(a) Resistivity versus temperature of Ca-214 single crystal. The $\rho(T)$
curves for $J=22$ (green) and $440$ (red) $\unit{mA/cm^{2}}$ measured
by first decreasing $T$ (continuous lines) and then heating the sample
(dashed lines) are highlighted. The blue triangles indicate when the
irreversibility described in the text sets in. Inset: selection of
$\rho(T)$ curves (labeled as a, b, c) plotted together with the fits
corresponding to different conduction regimes (VRH, SE and MS). (b)
$V-I$ characteristics as a function of $T$ on a double logarithmic
scale. A representative curve is labeled by the letter d. The black
solid circles connect the maximum of all the curves, $(V_{max},I_{max})$,
as better shown on linear scales (left inset). Right inset: temperature
dependence of $E_{max}$ (left scale) and $J_{max}$ (right scale).\label{fig2}}
\end{figure*}

X-Ray diffraction measurements in a specular $\omega$-$2\theta$
geometry ($\omega$ is the radiation incident angle on the sample
surface, while $2\theta$ is the angle between the incident and the
diffracted beam) were performed by using a Philips X'Pert-MRD high
resolution analytic diffractometer equipped with a four-circle cradle.
A $Cu$ $K_{\alpha1}$ ($\lambda=\unit[1.5406]{\mathring{A}}$) source
was used at $\unit[40]{kV}$ and $\unit[40]{mA}$. Measurements were
carried out by using monochromatic radiation obtained by equipping
the diffractometer with a four crystal Ge 220 Bartels asymmetric monochromator
and a graded parabolic Guttman mirror positioned on the primary arm.
On the secondary arm, the diffracted beam reaches the detector with
an angular divergence of 12 arcsec crossing a triple axis attachment
and undergoing three $(002)$ reflections within a channel cut Ge
Crystal.

\section{Results}

\subsection{Electrical transport measurements}

The temperature dependence of the resistivity measured along the \emph{c}
axis for selected values of $J$ is reported in semi-logarithmic scale
in Fig.~\ref{fig2}(a). It is important to notice that analogous
results were obtained on all the investigated samples. By increasing
$J$, $\rho$ is lowered of up to four orders of magnitude \citep{Okazaki_13,Sow_17}.
Moreover, despite $\rho$ is always a decreasing function of the temperature
($d\rho/dT<0$) \citep{Cao_97}, the shape of the resistivity curves
evolves as $J$ is increased and distinct $\rho(T)$ behaviors can
be observed, as indicated in the inset of Fig.~\ref{fig2}(a) by
the labels VRH, SE, and MS, which stand for Variable Range Hopping,
Semiconducting and Metastable, respectively, as discussed more in
detail in the following. In addition, a critical value $J^{sep}\approx\unit[0.4]{mA/cm^{2}}$
can be identified, which sets the change in the concavity of the $\rho(T)$
curves in semi-logarithmic scale, in accordance with Ref. \citep{Sow_17}.
The curves measured for $J<J^{sep}$ hardly depend on the value of
$J$, as in the case of the ones for $J=0.2$ and $\unit[0.4]{mA/cm^{2}}$,
which completely overlap \citep{Okazaki_13}. By measuring $\rho$,
both lowering and increasing $T$, an irreversible behavior, never
reported in the literature, was observed. Indeed, there are portions
of the $\rho(T)$ curves whose accessibility depends on the sample
history, as shown for example for $J=22$ and $\unit[440]{mA/cm^{2}}$.
Here the continuous lines indicate the data obtained by lowering $T$.
For $J<J^{sep}$, the resistance surge beyond the measurable range
of the voltmeter at a characteristic temperature, $T^{irr}\approx\unit[130]{K}$,
while for $J>J^{sep}$ the resistance is still measurable below this
value. However, by increasing the temperature from the lowest $T$
reached in the experiment, a measurable $\rho$ is detected only for
$T>T^{irr}$ (black dotted lines). The values of $T^{irr}$ are represented
as blue triangles in the figure. Interestingly, for all the analyzed
crystals and independently on $J$, $T^{irr}\approx\unit[130]{K}$,
a value comparable with $T_{AFM}$. This is the first time that a
measure of $\rho(T)$ gives indications of the magnetic ordering temperature
in Ca-214 \citep{Cao_97}. Moreover, this result confirms that $J$
induces a new more-conductive MS state where AFM is suppressed \citep{Bertinshaw_18},
and, more generally, that $J$ can be used to control the magnetic
ordering of this class of materials \citep{Sow_17,Bertinshaw_18,Sow_19}.
A more detailed analysis of this result is beyond the scope of this
work and will be subject of future studies.

In Fig.~\ref{fig2}(b), a selection of $V-I$ characteristics as
a function of $T$ obtained by $I$ biasing the sample along the \emph{c}
axis is shown on a double logarithmic scale. Beyond the low $J$ regime,
when the samples show a clear insulating behavior, a negative differential
resistance is observed \citep{Sakaki_13,Zhang_19}, in accordance
with the dramatic reduction of resistivity observed in the $\rho(T)$
curves by increasing $J$. By further increasing the current, an ohmic
dependence, signature of the IMT, is expected \citep{Nakamura_13}.
However, this threshold was not exceeded to preserve the crystal integrity
and measure the whole resistivity map on the same sample. The change
in the conduction results in a maximum in the characteristics at $(V_{max},I_{max})$
{[}or equivalently at $(E_{max},J_{max})${]}, as highlighted in Fig.~\ref{fig2}(b)
by black circles, both in the main panel and in the inset on the left,
where the shape of the curves on a linear scale can be better appreciated.
At room temperature $E_{max}\approx\unit[100]{V/cm}$ and $J_{max}\approx\unit[0.9]{A/cm^{2}}$.
Their temperature dependence is reported in the right inset of Fig.~\ref{fig2}(b).
While $E_{max}$ (black points, left scale) increases with $T$ \citep{Nakamura_13},
$J_{max}$ (red points, right scale) decreases on cooling. This latter
behavior is counter-intuitive and requires further analysis to be
understood. It is worth noting that $E_{max}$ should not be confused
with $E_{th}$. $E_{th}$ is the value at which, driving with electrical
potential, one reaches the thermodynamic metallic phase (M-long) \citep{Nakamura_13},
while $E_{max}$ is the value at which, driving with $J$, one reaches
the MS state.

\begin{figure}
\noindent \begin{centering}
\begin{tabular}{c}
\includegraphics[width=8cm]{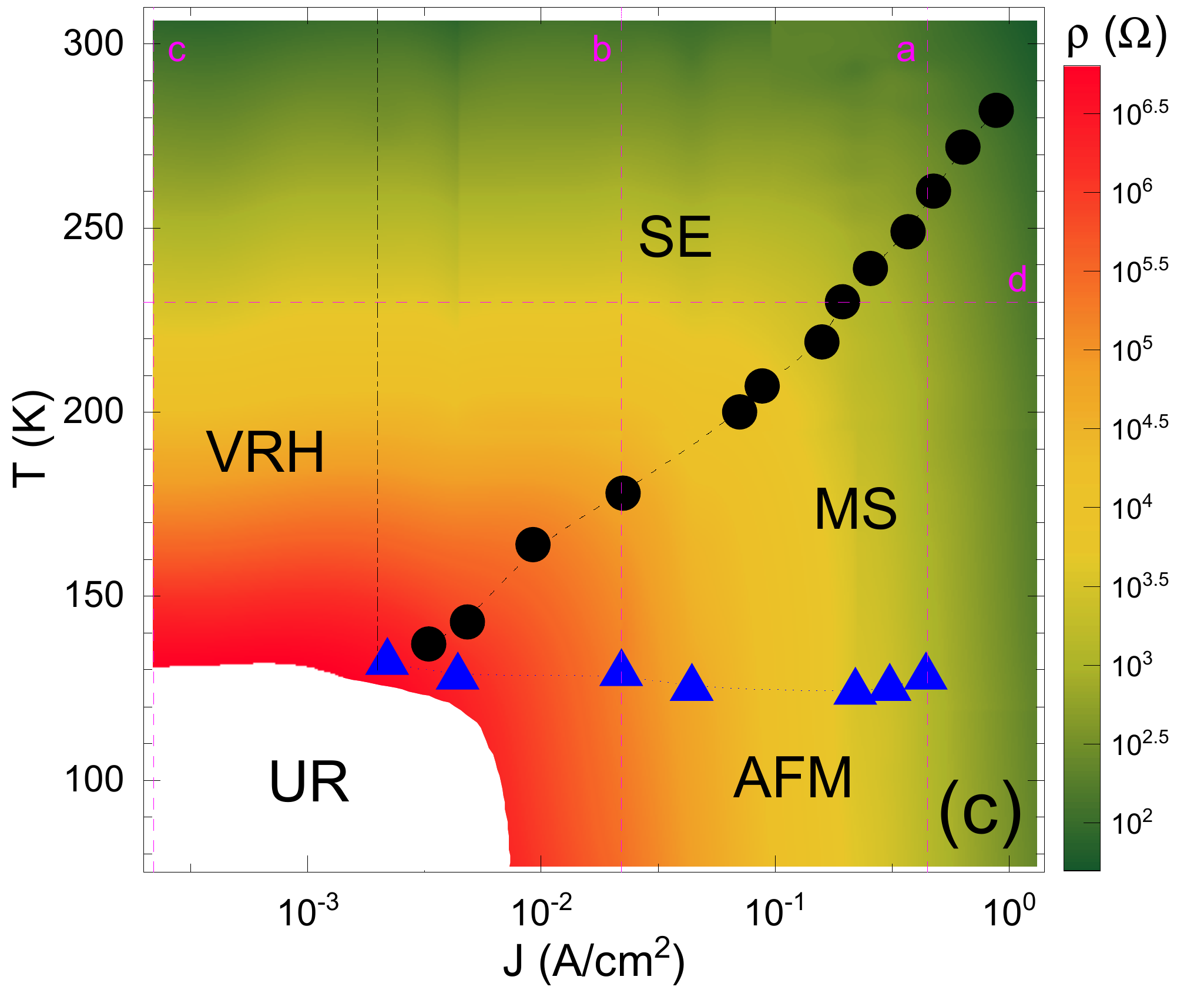}\tabularnewline
\end{tabular}
\par\end{centering}
\caption{$\rho(J,T)$ contour plot of the crystal resistivity obtained by combining
the $\rho(T)$ and the $V-I$ curves. The $\rho(T)$ {[}($V-I$){]}
curves labeled by the letter a, b and c (d) in Fig. \ref{fig2}(a)
{[}(b){]} are reported here. The different regions corresponding to
different conducting regimes (UR, VRH, SE, MS, and AFM) are highlighted.
The blue triangles (black circles) are the same of panel (a) {[}(b){]}
of Fig. \ref{fig2}.\label{fig3}}
\end{figure}

By combining both $\rho(T)$ curves measured for different values
of $J$ and $V-I$ characteristics as a function of $T$, it is possible
to draw the $\rho(J,T)$ contour plot of the crystal resistivity shown
in Fig.~\ref{fig3}. For the sake of clarity, only a selection of
$\rho(T)$ curves, representative of different conduction behaviors,
are reported in the Figure as vertical lines (a, b, c), while the
same $V-I$ curve labeled as (d) in Fig.~\ref{fig2}(b) is represented
as an horizontal line (see Fig.~\ref{fig3}). The resulting $\rho(J,T)$
phase diagram comprises different regions, corresponding to different
conducting regimes (UR, VRH, SE, MS, and AFM), as marked by the three
contours present in the figure. The dot-dashed vertical line represents
the value of $J^{sep}$. The position of the maximum of the $V-I$
curves at the investigated temperature are represented by black dashes
{[}as in Fig.~\ref{fig2}(b){]}. Finally, the blue dotted line at
$T^{irr}\approx\unit[130]{K}$ indicates the non-reversible behavior
of the $\rho(T)$ curves, namely the onset of the AFM ordering.

Accordingly, the following conducting regimes are identified. First,
in the limit of both low $J$ and $T$, there is the so-called Unexplored
Region (UR), namely a deeply insulating region which is not accessible
due to the limit of used experimental set-up. Then, by moving along
the $J$ axis ($J<J^{sep}$, all $T$), the $\rho(T)$ has a Variable
Range Hopping behavior with a power coefficient of about $0.25$,
typical of 3D systems (for all the details about the fitting of the
$\rho(T)$ curves the reader can refer to the Appendix). Here the
resistivity is not affected by the bias current density. For $J>J^{sep}$,
namely by crossing the dot-dashed line, a reduction of $\rho$ is
observed \citep{Okazaki_13}. From this side, regions SE and MS, divided
by the black dashed line, identify, respectively, the semiconducting
and the meta-stable regimes. In region SE ($J_{max}>J>J^{sep}$, $T>T_{max}$),
the best $\rho(T)$ fit is obtained by using a decreasing (negative)
exponential behavior resembling that of an intrinsic semiconductor
at sufficiently high $T$, that is a shallow insulator whose gap is
comparable to the temperature range under analysis. In region MS ($J>J_{max}$,
$T<T_{max}$) the $\rho(T)$ has a behavior that is very different
from both that of an insulator (decreasing, positive concavity in
both linear and log scale) and of a metal (increasing, positive concavity
in both linear and log scale), but a decreasing behavior with negative
(positive) concavity in log (normal) scale is measured. Indeed, this
change of concavity in the log scale allows to identify $J^{sep}$.
Such a situation, still interpreted in the VRH paradigm, marks the
divergency of the localization length. This can be interpreted as
the signal that at least a portion of the system becomes conducting,
leading to a resistivity that strongly resembles those of alloys and
whose best fit is obtained with a decreasing (negative) exponential
with a power coefficient of about 3. Which is the exact type of conducting
mechanism remains to be investigated. The intrinsic dependence on
time of the process makes difficult to characterize it through instruments,
and related concepts, that are meant to work at equilibrium.

\subsection{X-ray diffraction measurements}

\begin{figure}
\noindent \begin{centering}
\includegraphics[width=8cm]{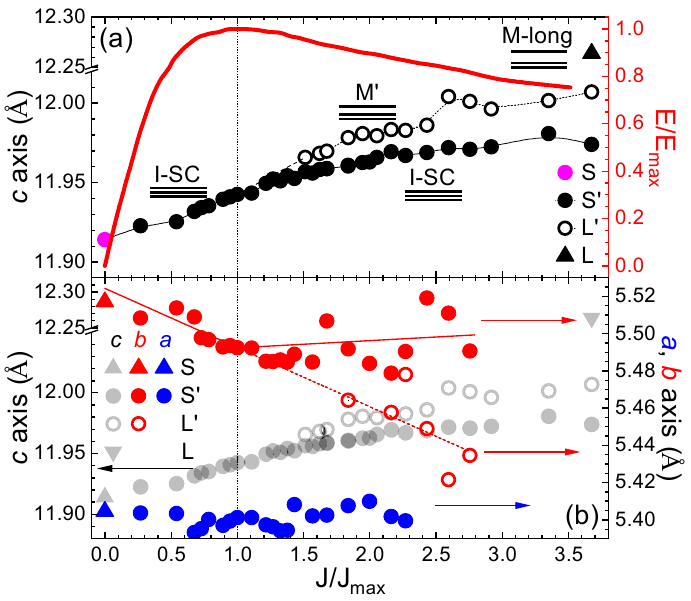}
\par\end{centering}
\noindent \caption{(a) Left scale: dependence of the \textit{c-}lattice parameter corresponding
to the different crystallographic phases (S, S', L' and L) as a function
of $J/J_{max}$. Right scale: normalized $V-I$ curve, $E/E_{max}-J/J_{max}$,
measured on the same crystal. Characteristic level arrangements (I-SC,
M', M-long) are also reported. (b) On the left (right) scale the dependence
of the \emph{c-} (\emph{a}-,\emph{ b}-) lattice parameter as a function
of $J/J_{max}$ is plotted. The error bars are smaller than the data
symbols. Lines are guides to eye. \label{fig4}}
\end{figure}

XRD measurements were performed at room $T$ by current biasing the
crystal to complement the electronic characterization and gain access
to the microscopic properties of the different conducting regimes.
In Fig.~\ref{fig4}(a), the dependence of the \textit{c}-lattice
parameter (left scale) on the normalized current density, $J/J_{max}$,
is superimposed to the normalized $E-J$ characteristic (right scale),
to allow the comparison among different samples; characteristic level
arrangements (I-SC, M', M-long, see below) are also reported. The
values of \textit{c} were calculated according to the Bragg law by
following the position of the $(006)$ reflection of the XRD $\omega$-$2\theta$
scans. The values of the \textit{c }\textit{\emph{axis}} plotted by
black-closed circles represent the elongation of the short \textit{c
}\textit{\emph{axis}} of the insulating S phase at $J=0$ ($c=\unit[11.914]{\mathring{A}}$,
magenta-closed circle) \citep{Nakamura_13}. This change produces
a distortion of the lattice cell, which now is labeled as S'. Interestingly,
at $J\approx J_{max}$ a new phase indicated as L', and represented
by open circles, clearly emerges. The \textit{c }\textit{\emph{axis}}
of L' also elongates by increasing $J$ and is well detectable in
the whole investigated current range, which covers the region of negative
differential resistance of the $E-J$ curve. Finally, at $J/J_{max}\approx3.7$,
the diffraction peak associated with the metallic L phase appears
($c=\unit[12.264]{\mathring{A}}$, black triangle) \citep{Nakamura_13}.

These measurements demonstrate that in the MS state, a new, possibly
metallic (L'), crystallographic phase coexists with the short insulating
one (S') in a quite wide range of current values and even with the
metallic one (L), at the maximum current reached in the experiment.
In panel (b), the dependence on $J/J_{max}$of the lattice parameters
\textit{a} and \textit{b, }calculated from the position of the reflections
$(208)$ and $(028)$ respectively, is compared with the \textit{c
}\textit{\emph{axis}}. Noticeably, while the value of the \textit{a}-axis
is almost constant, the \textit{b}-axis (red dots) splits in two branches
at $J/J_{max}\approx1$ as the \textit{c}-axis does. In the same region,
corresponding to the appearance of the L' phase, statistical scattering
is present in the \textit{b}-axis data. This can be interpreted as
a tentative of the system to release the in-plane strain while trying
to accommodate both phases (S' and L') in the crystal. From a careful
inspection of the data, it also emerges that the statistical scattering
of the two phases result overall complementary. It is worth noting
that the L' phase moves towards a metallic tetragonal structure, while
the S' phase slowly relaxes back towards the S one (in terms of crystallographic
axis). Indeed, once the L' phase nucleates and develops, S will sustain
only a smaller fraction of the flowing current. It is worth noticing
that the values of the lattice parameters, both in the S and in the
L phases, are consistent with the results reported in the literature
\citep{Braden_98,Alexander_99,Nakamura_13}. In particular, the value
of the \textit{c} axis in the metallic phase (L) is in accordance
with the ones reported for structural transitions induced by electric
field, pressure and temperature \citep{Nakamura_13}. This indicates
that, contrary to the MS state, L is a real thermodynamic phase.

\section{Discussion}

The emergence of a metallic phase (L') in the system would explain
both the puzzling negative differential resistivity of the MS regime
and the counter-intuitive increase of $J_{max}$ with $T$. In fact,
in order to sustain a systematic increase of current flow in an overall
insulating state, at a certain critical current density, dependent
on temperature, and comparable with $J_{max}(T)$, the system finds
energetically more convenient to nucleate a more conductive crystallographic
phase, L'. Consequently, above $J_{max}(T)$, the electrical potential
needed to further increase the current flow reduces, while the more-conductive
L' phase grows. On increasing $T$, the S' phase itself can sustain
more current, since it becomes less insulating. Accordingly, $J_{max}(T)$
is an increasing function of $T$. This is just one of the clearest
signatures that the emergence of L' is not a classical effect driven
by Joule heating, but that it comes from a much more subtle and complex
energy balance.

The remarkable increase of \textit{c} and decrease of \textit{b} in
L' phase is definitely compatible with a significant decrease of the
ratio $\bar{x}/z$ that would steadily lead to a metallic behavior
of that portion of the material. To check this hypothesis, a transformation
matrix computed in Ref. \citep{Han_18} by means of DFT+U calculations
was used. This allows to track the effect of applied strain on \textit{a,
b} and \textit{c} and, in particular, as this reflects on $x$, $y$
and $z$ (see Appendix). The obtained related changes of $x$, $y$
and $z$ give, as expected, a decreasing ratio of $\bar{x}/z$, following
the evolution with increasing current from S' to L', but also two
unpredictable results: first, above $J_{max}$, that is, once L' sets
in, S' goes back towards the values of $x$, $y$ and $z$ characteristics
of S; second, the decrease of $\bar{x}/z$ is mainly determined by
the decrease of $x$ and not by the increase of $z$. Once the system
has the possibility to fully exploit the L' phase to allow an increasingly
current flow, S' phase can relax back to S one. The complicated intertwining
of rotation, tilt and distortion maps the increase of \textit{c} mainly
on a reduction of $\bar{x}$ than on an increase of $z$.

\section{Conclusions}

In summary, dc current drive was used to determine the $\rho(J,T)$
phase diagram. By profiting of a new protocol, it was possible to
access a region of the phase diagram not yet explored and to unveil
the nucleation and evolution of a new metallic crystallographic phase,
L', completely compatible with the transport data. Its corresponding
cell dimensions depart from those of the insulating \emph{short} phase
and approach those of the metallic \emph{long} phase. The main octahedral
axis and the corresponding $Ru$ levels of the new phase were theoretically
obtained: the phase L' is more conducting than S' and can be considered
as a precursor of the metallic L phase.

Such findings explain the unexpected and counterintuitive results
of the transport data and completely determine the behavior in the
metastable phase. Such findings are consistent with the literature,
and represent a significant improvement of the current comprehension
of a complex system such as Ca-214, opening new perspectives in its
microscopic characterization. These results open new perspectives
in the microscopic characterization of Ca-214. For instance, spectroscopic
measurements under electrical current drive may represent a valuable
validation of the present findings.
\begin{acknowledgments}
The authors gratefully acknowledge Y. Maeno and G. Mattoni for the
useful discussions and I. Nunziata for technical support.
\end{acknowledgments}

\appendix

\section{XRD data supplement}

\begin{figure}
\noindent \centering{}\includegraphics[width=8cm]{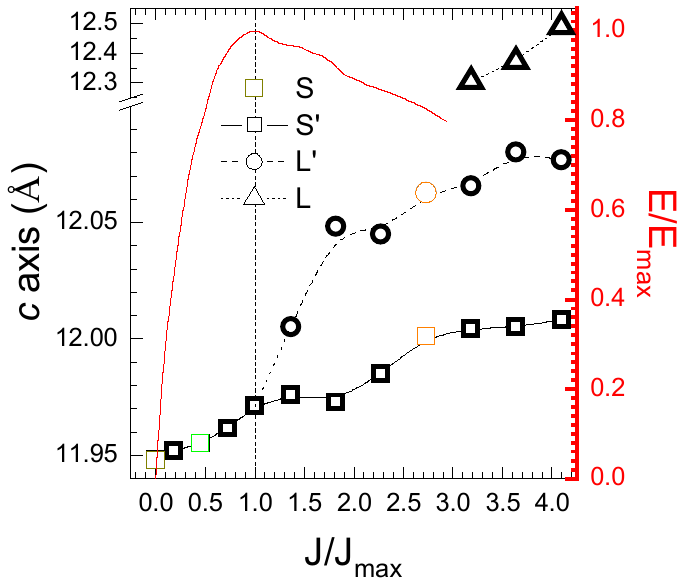}\caption{Left scale: dependence of the \textit{c} lattice parameter corresponding
to the different crystallographic phases (S, S', L', and L) at room
temperature as a function of the normalized bias current density $J/J_{max}$.
The bigger colored points indicate the values of the \textit{c }\textit{\emph{axis}}
extracted from the XRD scans of the corresponding colour reported
in Fig. 2. Right scale: normalized $V-I$ curve, $E/E_{max}-J/J_{max}$,
measured at room temperature on the same crystal.\label{fig.S1}}
\end{figure}

\begin{figure}
\noindent \centering{}\includegraphics[width=8cm]{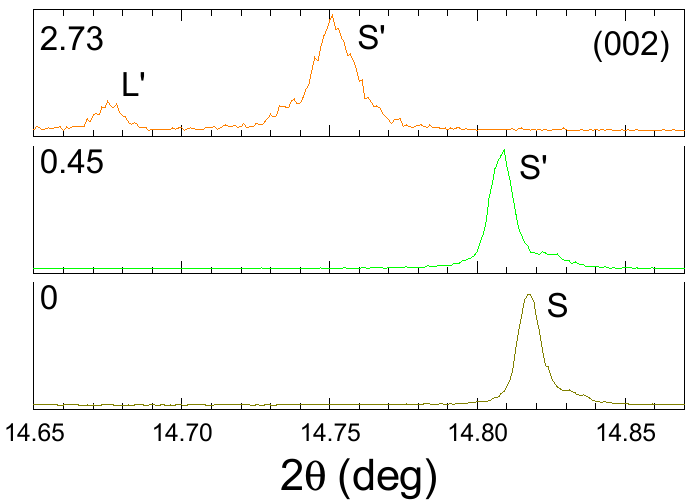}\caption{Representative $\omega$-$2\theta$ scans of the $(002)$ reflections
are reported for different values of the normalized bias current $J/J_{max}$
for the same crystal of Fig. ~\ref{fig.S1}. The labels on the diffraction
peaks correspond to the different crystallographic phases (S, S',
L') present in the different conduction regimes. \label{fig.S2}}
\end{figure}

In order to provide further evidence of the coexistence of the three
distinct crystallographic phases in the current induced meta-stable
state, additional XRD data for another single crystal are presented
in Fig.~\ref{fig.S1}. Here the values of the \textit{c}-lattice
parameters as a function of the normalized electrical current density
were derived from the $(002)$ reflection. Again the comparison with
the normalized $E-J$ curve measured for the same crystal (right scale)
confirms that the S' phase splits into the L' phase at $J\approx J_{max}=0.8$
$A/cm^{2}$ (see vertical dashed line). This new phase is well distinguishable
in all the investigated current range from the other two diffraction
peaks, as shown in Fig.~\ref{fig.S2}, where three representative
$\omega$-$2\theta$ scans of the $(002)$ reflection are reported
for different values of $J/J_{max}$ corresponding to different regions
of the $E/E_{max}-J/J_{max}$ characteristic. It is evident that before
reaching the maximum of the $E/E_{max}-J/J_{max}$ characteristic,
namely in the insulating regime, only the peaks identifying the phases
S and S' are present (dark yellow and green scans, respectively).
Above $J_{max}$, the diffraction peak of the L' phase develops, as
shown by the orange line, acquired at $J/J_{max}=2.73$.

\section{Theoretical methods}

\begin{figure*}
\noindent \begin{centering}
\begin{tabular}{ccc}
\includegraphics[width=5.3cm]{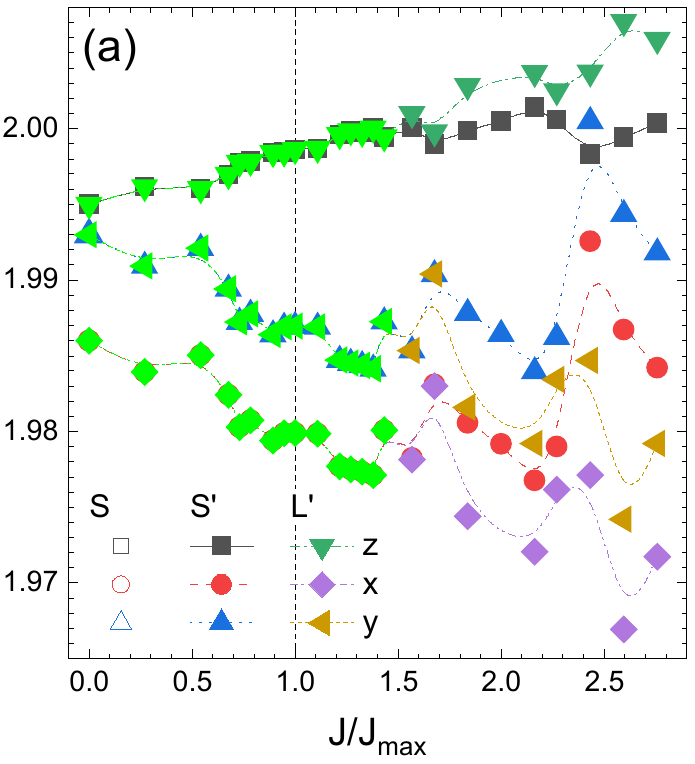}  & \includegraphics[width=5.3cm]{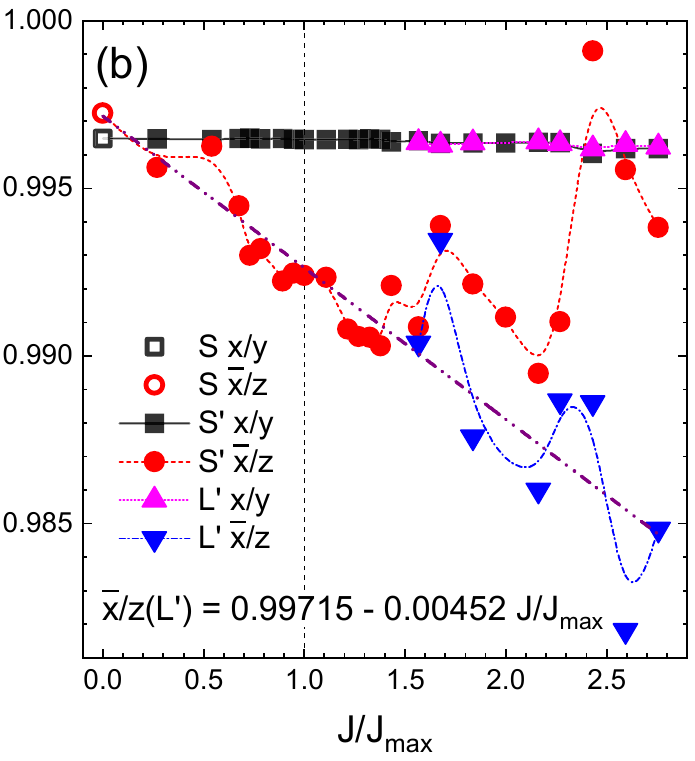}  & \includegraphics[width=5.6cm]{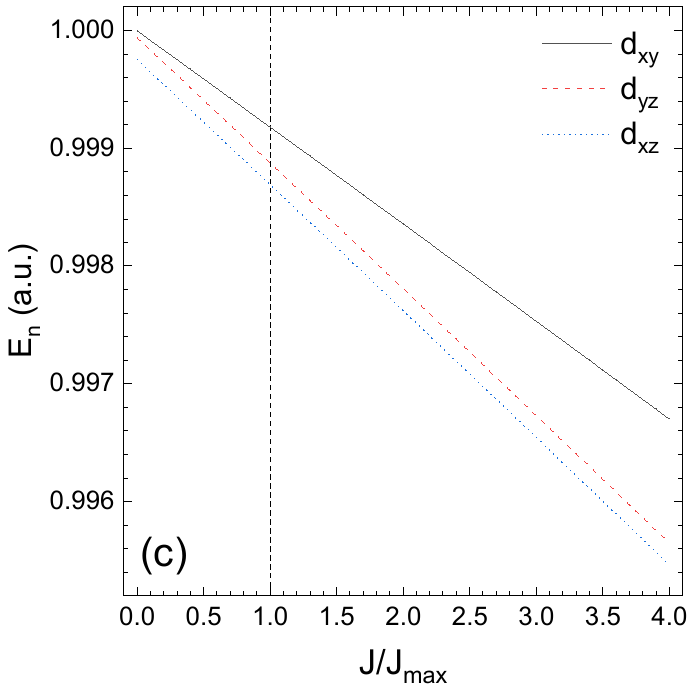}\tabularnewline
\end{tabular}
\par\end{centering}
\caption{$x$, $y$ and $z$ (a) and $x/y$ and $\bar{x}/z$ (b) as functions
of $J/J_{max}$ for the phases S, S' and L' . (c) Energies (in arbitrary
units) of the $Ru$ $d_{xy}$, $d_{yz}$ and $d_{xz}$ $RuO_{6}$
octahedra levels as functions of $J/J_{max}$ for L' phase (S' phase
for $J\protect\leq J_{max}$).\label{fig.S0}}
\end{figure*}

\subsection{$RuO_{6}$ octahedra}

\begin{figure*}
\noindent \begin{centering}
\begin{tabular}{cccc}
\includegraphics[width=4.5cm]{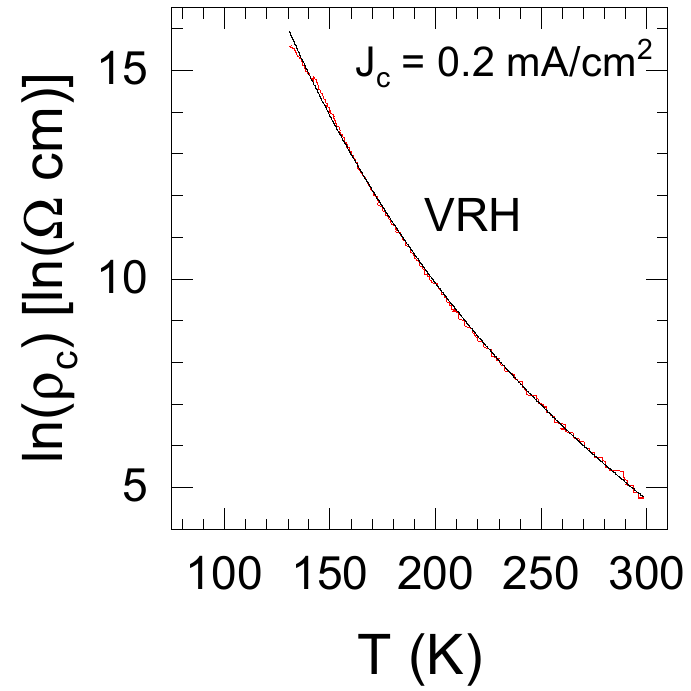}  & \includegraphics[width=4.5cm]{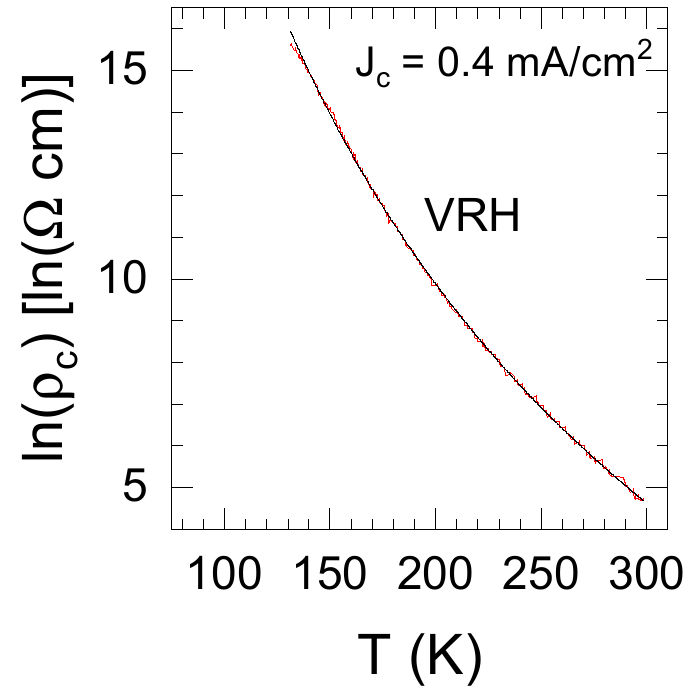}  & \includegraphics[width=4.5cm]{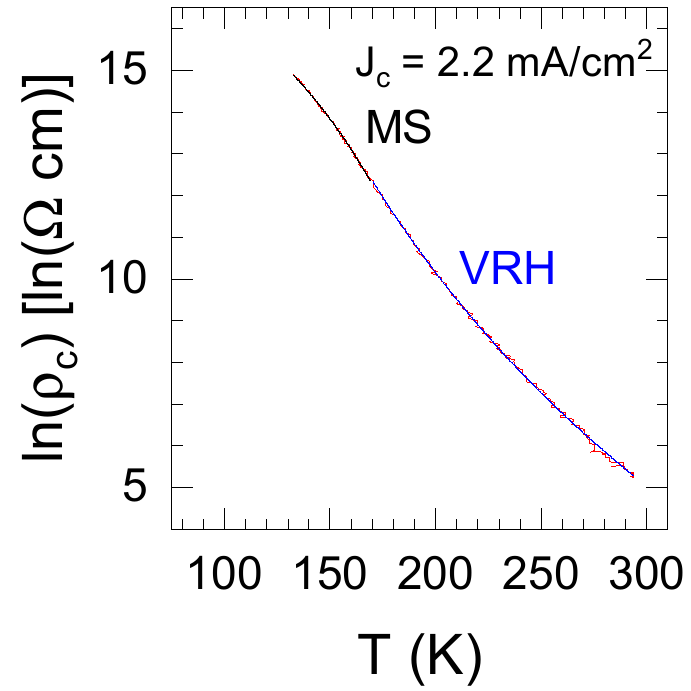}  & \includegraphics[width=4.5cm]{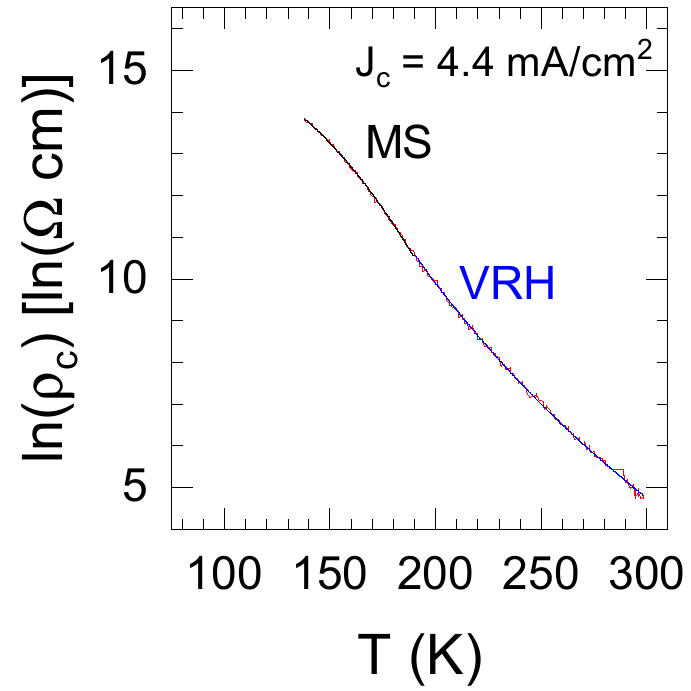}\tabularnewline
\includegraphics[width=4.5cm]{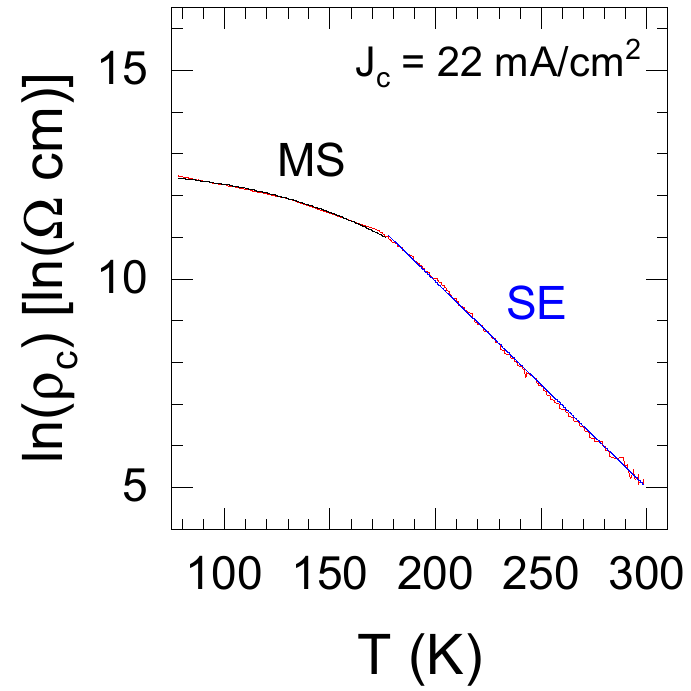}  & \includegraphics[width=4.5cm]{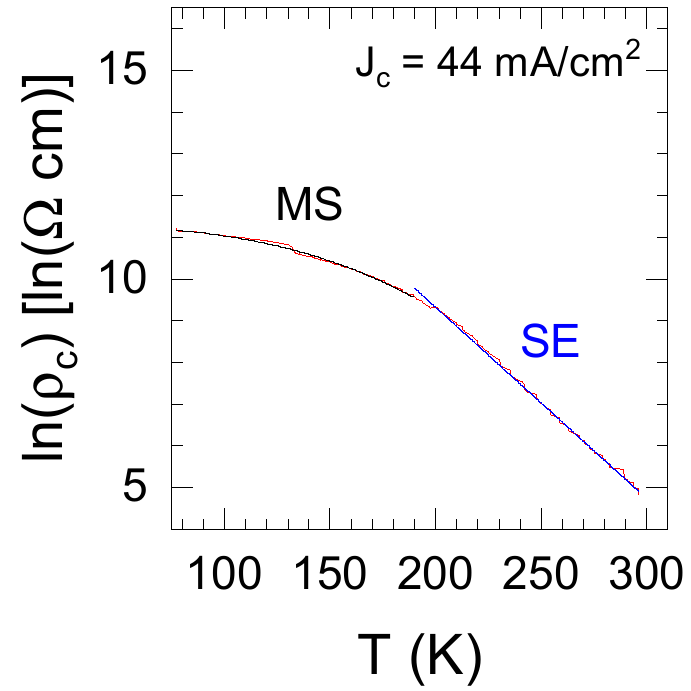}  & \includegraphics[width=4.5cm]{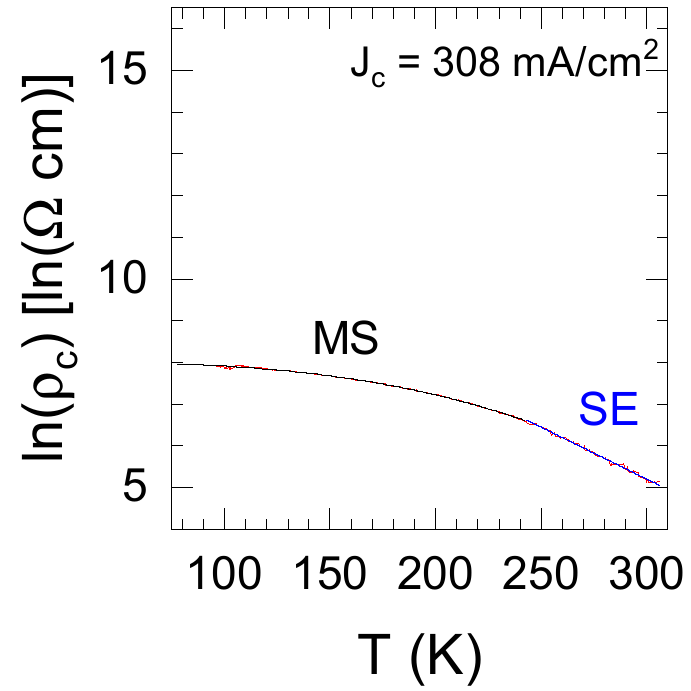}  & \includegraphics[width=4.5cm]{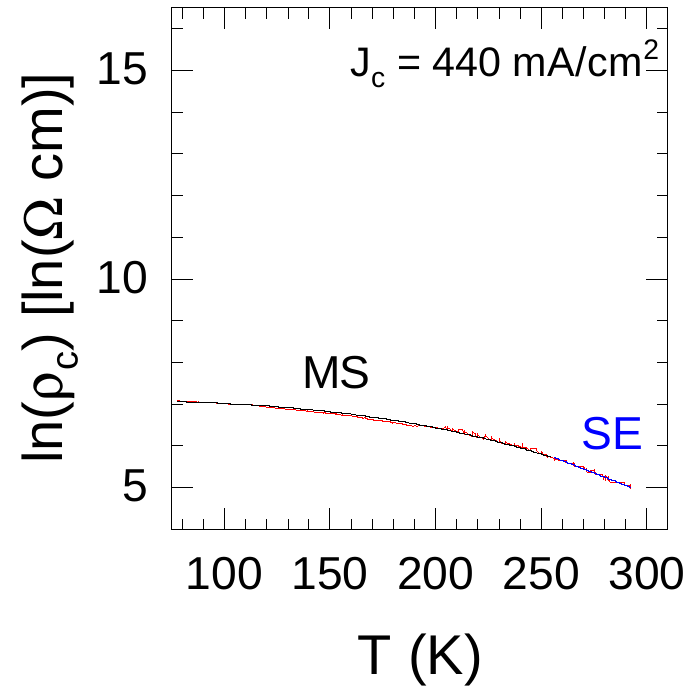}\tabularnewline
\end{tabular}
\par\end{centering}
\caption{Least-squares fits of $\ln\left(\rho_{c}\right)$ as a function of
$T$ for various values of $J_{c}$.\label{fig:fits}}
\end{figure*}

\subsubsection{Crystal field}

The $RuO_{6}$ complex is an octahedron whose vertices are occupied
by $6$ $O$ atoms and its center by a $Ru$ atom. Such a type of
$Ru$-$O$ coordination, according to the Jahn-Teller effect \citep{Khomskii_14},
splits the $d$ levels of the $Ru$ in two groups: $e_{g}$, $d_{x^{2}-y^{2}}$
and $d_{3z^{2}-r^{2}}$, and $t_{2g}$, $d_{xy}$, $d_{yz}$ and $d_{xz}$.
In the first group, $e_{g}$, the orbitals have lobes pointing directly
towards the directional $p$ orbitals of $O$ and therefore lie higher
in energy. On the other hand, in the second group, $t_{2g}$, the
actual distances of the apical oxygens $Ru-O2$, $z$ in the main
text, and of the in-plane oxygens $Ru-O1$, $x$ and $y$ in the main
text (and $\bar{x}$ their average), determine the degree of degeneracy
of the three levels: a perfect octahedron ($z=x=y$) leads to three
perfectly degenerate levels. Instead, the smaller is $x$ with respect
to $y$ (at fixed $z)$ the higher in energy lies the level $d_{xz}$
with respect to $d_{yz}$; as well as the smaller is $\bar{x}$ with
respect to $z$ the higher in energy lies the level $d_{xy}$ with
respect to the $d_{yz}$-$d_{xz}$ doublet.

As schematically reported in Fig. 1(b) in the main text, the order
in energy of the $t_{2g}$ levels is fundamental to establish how
the four electrons per $Ru$ present in the system decide to occupy
such levels. As a consequence, this determines the transport properties
of the related state. In the I-short state, $z/\bar{x}<1$ and $d_{xy}$
is lower in energy with respect to the $d_{yz}$-$d_{xz}$ doublet
with a crystal field gap that can be so large that the electrons prefer
to arrange in pairs in $d_{xy}$ level although the local Coulomb
repulsion would avoid that. The remaining two electrons can accommodate
in the $d_{yz}$-$d_{xz}$ doublet according to the Hund's rule with
parallel spins and such a configuration, at low enough temperatures,
leads to an insulating antiferromagnetic state. At higher temperatures,
since $z/\bar{x}$ gets closer and closer to $1$ the levels become
almost degenerate. In this situation, the strong correlations prevent
the system to behave as a metal, but still as an insulator, by splitting
the $d_{yz}$ and $d_{xz}$ levels in lower and upper Mott-Hubbard
bands. By further increasing the temperatures, $z/\bar{x}$ become
sufficiently larger than $1$ to have the $d_{yz}$-$d_{xz}$ doublet
below the $d_{xy}$ level and lead to a metal. In this case, three
electrons fill in the levels according to the Hund's rule and one
electron gets free to move in the lattice.

\subsubsection{Crystallographic axes vs $RuO$ distances}

By means of DFT+U calculations, A. Millis and coworkers \citep{Han_18}
found a transformation matrix relating the variations of the crystallographic
axis $\delta a$, $\delta b$ and $\delta c$ to the variations of
the $Ru-O$ distances in the $RuO_{6}$ octahedra, that is, $\delta x$,
$\delta y$ and $\delta z$: 
\begin{equation}
\left(\begin{array}{c}
\delta x^{+}\\
\delta z\\
\delta x^{-}
\end{array}\right)=\left(\begin{array}{ccc}
0.3740 & -0.0053 & -0.0698\\
-0.0517 & 0.0746 & 0.0313\\
-0.0082 & -0.0059 & -0.0015
\end{array}\right)\left(\begin{array}{c}
\delta a^{+}\\
\delta c\\
\delta a^{-}
\end{array}\right)
\end{equation}
where $\delta x^{\pm}=\frac{1}{\sqrt{2}}\left(x\pm y\right)$, and
$\delta a^{\pm}=\frac{1}{\sqrt{2}}\left(a\pm b\right)$. This matrix
allows one to find the values of $x$, $y$, and $z$ given those
of $a$, $b$ and $c$ for the the two phases, S' and L', emerging
from the S one on applying an electrical current drive {[}see Fig.~\ref{fig.S0}(a){]}.
It was then possible to obtain the two fundamental ratios $x/y$ and
$\bar{x}/z$ in the S' and L' phases {[}see Fig.~\ref{fig.S0}(b){]}.
A least-squares linear fit of the ratio $\bar{x}/z$ for the L' phase
(following the one of the S' phase for $J\leq J_{max}$) resulted
very accurate and the related fit parameters are reported directly
in the figure {[}see Fig.~\ref{fig.S0}(b){]}. Given the almost constant
ratio $x/y$ and the linear fit of the ratio $\bar{x}/z$, it has
been possible to compute the relative energies of the $d_{xy}$, $d_{yz}$
and $d_{xz}$ levels {[}see Fig.~\ref{fig.S0}(c){]}. This supports
our interpretation that the unconventional and puzzling behavior of
the meta-stable state are due to the emergence of the metallic phase
L' in the system.

\subsection{Conductive regimes: VRH, SE and MS}

All the curves reporting the behavior of the resistivity $\rho$ as
a function of the temperature $T$, for different values of $J$,
have been least-squares fitted with the same generic allometric function
(see Fig.~\ref{fig:fits}): 
\begin{equation}
\ln\left(\rho\right)=A+BT^{C}
\end{equation}
According to the sign of $B$ and the value of $C$, it is possible
to identify three distinct conducting regimes (VHR, SE and MS) which
set in a specific range of temperatures depending on the value $J$
(see Tabs. \ref{tab:VRH}-\ref{tab:SE}). The values of $C$ have
been chosen according to the closest value for all currents and temperatures
in the regime in order to avoid excessive fluctuations in the other
parameters.

It is worth noting that such unbiased fits of the R(T) curves independently
and accurately reproduce the position of the maximum in the I-V characteristics.

\begin{table}
\noindent \begin{centering}
\begin{tabular}{|c|c|c|c|c|}
\hline 
$J$ ($\unit{mA/cm^{2}}$)  & $T$ ($\unit{K}$)  & A  & $B$  & \emph{$C$}\tabularnewline
\hline 
\hline 
$0.2$  & All  & $-43.9$  & $202.2$  & \emph{$-0.25$}\tabularnewline
\hline 
$0.4$  & All  & $-44.6$  & $205.1$  & \emph{$-0.25$}\tabularnewline
\hline 
$2.2$  & $>170$  & $-43.0$  & $199.7$  & \emph{$-0.25$}\tabularnewline
\hline 
$4.4$  & $>190$  & $-43.5$  & $200.7$  & \emph{$-0.25$}\tabularnewline
\hline 
\end{tabular}
\par\end{centering}
\caption{Variable Range Hopping regime fitting parameters\label{tab:VRH}}
\end{table}

\subsubsection{Variable Range Hopping (VRH)}

In this case, it is $B>0$ and $C<0$. The results of the fitting
procedure reported in Tab.~\ref{tab:VRH} are compatible with a 3D
system. 
\begin{quote}
\begin{table}
\noindent \begin{centering}
\begin{tabular}{|c|c|c|c|c|c|}
\hline 
$J$ ($\unit{mA/cm^{2}}$)  & $T$ ($\unit{K}$)  & A  & $B$  & $C$  & $T_{0}$ ($\unit{K}$)\tabularnewline
\hline 
\hline 
$2.2$  & $<170$  & $17.2$  & $-1.01\times10^{-6}$  & $3$  & $99.6$\tabularnewline
\hline 
$4.4$  & $<190$  & $15.9$  & $-7.89\times10^{-7}$  & $3$  & $108$\tabularnewline
\hline 
$22$  & $<177$  & $12.5$  & $-2.82\times10^{-7}$  & $3$  & $152$\tabularnewline
\hline 
$44$  & $<189$  & $11.3$  & $-2.53\times10^{-7}$  & $3$  & $158$\tabularnewline
\hline 
$308$  & $<243$  & $8.01$  & $-9.84\times10^{-8}$  & $3$  & $217$\tabularnewline
\hline 
$440$  & $<256$  & $7.10$  & $-8.31\times10^{-8}$  & $3$  & $229$\tabularnewline
\hline 
\end{tabular}
\par\end{centering}
\caption{Meta-Stable regime fitting parameters\label{tab:MS}}
\end{table}
\end{quote}

\subsubsection{Meta-Stable (MS)}

In this case, it is $B<0$ and $C>0$. In Tab.~\ref{tab:MS} the
fitting parameters corresponding to the MS regime are reported. $T_{0}=\left|B\right|^{-\frac{1}{C}}$
is the equivalent activation temperature. 
\begin{quotation}
\begin{table}
\noindent \begin{centering}
\begin{tabular}{|c|c|c|c|c|c|}
\hline 
$J$ ($\unit{mA/cm^{2}}$)  & $T$ ($\unit{K}$)  & $A$  &  & $ $  & $T_{0}$ ($\unit{K}$)\tabularnewline
\hline 
\hline 
$22$  & $>177$  & $19.8$  & $-0.0494$  & $1$  & $20.2$\tabularnewline
\hline 
$44$  & $>189$  & $18.5$  & $-0.0458$  & $1$  & $21.8$\tabularnewline
\hline 
$308$  & $>243$  & $12.7$  & $-0.0249$  & $1$  & $40.2$\tabularnewline
\hline 
$440$  & $>256$  & $10.8$  & $-0.0197$  & $1$  & $50.8$\tabularnewline
\hline 
\end{tabular}
\par\end{centering}
\caption{Semiconductor regime fitting parameters\label{tab:SE}}
\end{table}
\end{quotation}

\subsubsection{Semiconductor (SE)}

In this case, it is $B<0$ and $C=1$. The fitting procedure returns
the values reported in Tab.~\ref{tab:SE}. $T_{0}=\left|B\right|^{-1}$
is the activation temperature.

%


\end{document}